\documentclass{fundam}

\usepackage[english]{babel}
\usepackage[T1]{fontenc}
\usepackage[utf8]{inputenc}
\usepackage{microtype}

\usepackage{amsmath}
\usepackage{amssymb}
\usepackage{graphicx}
\usepackage{pgf}
\usepackage{tikz}
\usetikzlibrary{arrows,automata}
\usepackage{booktabs}
\usepackage{xspace}

\newcommand{\Z}{\mathbb{Z}}

\newcommand{\brsim}{\texttt{brsim}\xspace}
\newcommand{\heresy}{\texttt{HERESY}\xspace}
\newcommand{\clrs}{\texttt{cl-rs}\xspace}
\newcommand{\rsA}{\mathcal{A}\xspace}

\newcommand{\matR}{\mathbf{R}}
\newcommand{\matI}{\mathbf{In}}
\newcommand{\matP}{\mathbf{P}}

\DeclareMathOperator{\res}{res}

\let\vec\mathbf

\title{The Many Roads to the Simulation of Reaction Systems}

\author{Claudio Ferretti \and
  Alberto Leporati \and
  Luca Manzoni\\
  Dipartimento di Informatica, Sistemistica e Comunicazione\\
  Università degli Studi di Milano-Bicocca\\
  Viale Sarca 336, 20126 Milan, Italy\\
  \{ferretti, leporati, luca.manzoni\}@disco.unimib.it
  \and
  Antonio E. Porreca \\
  Aix Marseille Université, Université de Toulon, CNRS, LIS, Marseille, France \\
  antonio.porreca@lis-lab.fr}

\runninghead{C. Ferretti et Al.}{The Many Roads to the Simulation of Reaction Systems}

\begin{document}

\maketitle

\begin{abstract}
  Reaction systems are a computational model inspired by the bio-chemical reactions that happen inside biological cells. They have been and currently are studied for their many nice theoretical properties. They are also a useful modeling tool for biochemical systems, but in order to be able to employ them effectively in the field the presence of efficient and widely available simulators is essential. Here we explore three different algorithms and implementations of the simulation, comparing them to the current state of the art. We also show that we can obtain performances comparable to GPU-based simulations on real-world systems by using a carefully tuned CPU-based simulator.
\end{abstract}


\section{Introduction}
\label{sec:intro}

Reaction Systems are a novel and growing formalism based on the idea of biochemical reaction~\cite{Ehrenfeucht2005a,Ehrenfeucht2007a}. They are amenable to both theoretical studies and as a modeling tool for biological processes. As a computational model, they (and their dynamics) occupy an interesting intermediate position between Boolean Automata Networks~\cite{demongeot2010a,demongeot2012a} and Cellular Automata~\cite{Kari2005a,Dennunzio2013a}. The theoretical exploration is flourishing, with the investigation of combinatorial properties~\cite{Ehrenfeucht2010a,Dennunzio2015c}, complexity of establishing the presence of dynamical behaviours~\cite{Dennunzio2015a,Azimi2016a,DennunzioEtAl_IC2019}, causal dynamics~\cite{Barbuti2015a,Barbuti2016a,Barbuti2016b}, and  the classification of reaction systems according to the relation of mutual simulability~\cite{Manzoni2013b}. Reaction systems, however, have also been employed to model real-world systems~\cite{Corolli2012,Azimi2014b}. The availability of fast and efficient simulators is essential for a more widespread use of reaction systems as a modeling tool. The first widely available simulator was \brsim~\cite{azimi2015dependency}, available at~\cite{brsim_github,brsim_webpage}. It is written in Haskell and, until now, it was the fastest CPU-based simulator available. Its development continues with the addition of nice user-friendly features, and the ability to explore more properties of reaction systems, not only for the simulation of the dynamics~\cite{ivanov2018a}.

Recently, a GPU-based approach to the simulation of reaction systems has been explored with \heresy~\cite{Nobile2017a}, available at~\cite{heresy_github}. The GPU-based simulator written using CUDA proved to be the fastest one for large-scale systems, due to its ability to exploit the large number of computational units inside GPUs. Even if \heresy also provides a CPU-based simulator written in Python 2, it is more a ``fallback'' simulator when GPUs are not available, and is slower than \brsim.

Both simulators, however, employ the same direct simulation method (i.e., based directly on the set-theoretic definition of the reaction systems' dynamics). Here we provide an optimized Common Lisp~\cite{steele1990common} simulator, called \clrs~\cite{clrs_github}, also employing the direct simulation method, which is able to offer performances comparable with the GPU-based simulator on a large-scale real-world model, the \emph{ErbB} model~\cite{helikar2013comprehensive}. \clrs proves to be the fastest CPU-based simulator currently available. We also explore other ways of performing the simulation, in particular:
\begin{itemize}
  \item By looking at the graph of dependencies between reactions (i.e., which reactions produce the reactants required by other reactions), it is possible to avoid performing the simulation of parts of the reactions that cannot produce any effect on its dynamics. This mode of operation is also available in \clrs.
  \item By rewriting the dynamical evolution of a reaction system in terms of matrix-vector multiplications, vector additions, and clipping operations, it is possible to exploit the existing high-performance linear algebra libraries to perform the simulation. A proof-of-concept implementation employing Python 3 and Numpy~\cite{numpy_webpage} is used in this paper.
\end{itemize}

The rest of the paper is organized as follows. In Section~\ref{sec:basic} we recall the basic notions on reaction systems, in Section~\ref{sec:simulation} we introduce three different algorithms to simulate the evolution of the system. We describe the experimental settings used to compare them to the state of the art in Section~\ref{sec:settings}. The results of the comparison are presented in Section~\ref{sec:results}, while future works and possible directions of research are detailed in Section~\ref{sec:conclusion}.

\section{Basic Notions}
\label{sec:basic}

In this section we briefly recall the basic notions of reactions, reaction systems, and interactive processes that were first introduced in~\cite{Ehrenfeucht2007a}.

A reaction is a triple $a = (R_a, I_a, P_a)$ of non-empty and finite sets with $R_a \cap I_a = \varnothing$. The sets are called \emph{reactants}, \emph{inhibitors}, and \emph{products}, respectively. If all three sets are subsets of the same finite set $S$, then the reaction $a$ is said to be a reaction \emph{over} the \emph{background set} $S$ and its elements are called \emph{entities}. Given a set $T \subseteq S$ of symbols, and one reaction $a = (R_a, I_a, P_a)$ over $S$, the reaction $a$ is said to be enabled in $T$ when $R_a \subseteq T$ and $I_a \cap T = \varnothing$. In other words, a reaction is enabled if all reactants are present in $T$ but none of the inhibitors is. If a reaction is enabled in $T$, then the results of $a$ on $T$, denoted by $\res_a(T)$, is the products set $P_a$. If $a$ is not enabled in $T$, then $\res_a(T) = \varnothing$.

A reaction system is a pair $\rsA = (S, A)$, where $A$ is a set of reactions over $S$, the background set. The notion of result function can be extended to the entire set $A$ of reactions of the reaction system $\rsA$ in the following way:
\begin{align*}
  & \res_{\rsA}(T) = \bigcup_{a \in A} \res_a(T) & \text{for all $T \subseteq S$}
\end{align*}
that is, $\res_\rsA$ is a function from $2^S$ to itself, that can then be used to define the discrete-time dynamical system $(2^S, \res_\rsA)$, where $2^S$ is the set of states and $\res_\rsA$ is the function mapping each state (an element of $2^S$) to the next state. This process can be iterated to obtain the \emph{orbit} for each state $T \subseteq S$:
\begin{align*}
  & \left(T, \res_\rsA(T), \res_\rsA^2(T), \res_\rsA^3(T), \ldots\right)
\end{align*}
The orbit of a state $T$ gives the evolution with respect to time (represented in discrete steps) of the set of entities $T$. For simulation purposes, this corresponds to exploring how a set of chemical species evolves with time \emph{without} any external interaction.

To introduce the possibility of interaction with an external environment, the notion of \emph{interactive process} and of \emph{context sequence} was introduced in~\cite{Ehrenfeucht2007a}. The main idea of a context sequence is to have new entities inserted between each time step. This can be used to model the interaction with other systems. A context sequence is then a (finite) sequence $\mathcal{C}  = \left( C_0, C_1, \ldots, C_{n-1} \right)$ of subsets of $S$. The dynamics of the system is then described by the sequence $\mathcal{D} = \left( D_0, D_1, \ldots, D_n \right)$ where $D_0 = \varnothing$ and $D_i = \res_\rsA(D_{i-1} \cup C_{i-1})$ for all $1 \le i < n$.

\begin{example}
  As a working example, in the rest of the paper we will use a reaction system $\rsA$ with a background set of $S = \{a, b, c, d\}$ and the following three reactions:
  \begin{align*}
    & r_1 = (\{a, b\}, \{c\}, \{a, b\}) & r_2 = (\{a\}, \{b, c\}, \{d\}) &&  r_3 = (\{d\}, \{c\}, \{b\})
  \end{align*}
  As a starting set, we will consider the state $T = \{b, d\}$. Therefore $\res_\rsA(T) = \{b\}$, since only reaction $r_3$ is enabled, and then $\res_\rsA^2(T) = \varnothing$, since no reaction is enabled in $\{b\}$. The context is not considered in this example, and hence we do not provide any context sequence.
\end{example}

\section{Reaction Systems Simulation Algorithms}
\label{sec:simulation}

In this section we recall the classical ``direct'' simulation method, together with two other possible approaches, one based on observing the dependencies between reactions, and one based on modeling each time step as a series of operations on vectors and matrices.

In the following we always suppose to have both $S$ and $A$ endowed with an arbitrary linear order, so that terms like ``the $i$-th reaction'' will be well-defined.

\subsection{Direct Simulation}

The direct simulation directly follows from the set-based definition of reaction systems. Each state is represented as a set, and a reaction actually acts on it. The difference between the simulators is mainly due to the data structure actually employed to represent sets and reactions.

In \clrs, we decided to employ a bit-set based representation. The current state is a bit vector of $|S|$ bits. This representation might not be very compact for large sets, but for the largest models used in our tests the size of $2000$ bits ($250$ bytes) is small enough to fit in most processor caches, resulting in a limited access to the slower main memory. Each reaction is represented as a structure with three vectors representing the sets $R_a$, $I_a$, and $P_a$, respectively, each of them containing a collection of entities specified as indices for the bit set that represents the state of the system. Each index is a fixed-length binary number that can fit into the machine registers. This means that to test if an entity (for example, a reactant) is present, it is sufficient to check if the corresponding index in the bit set representing the state of the system is set to one. If the need to simulate larger systems arises, it would be possible to represent sets in a more compact way, for example using compressed bitmaps, while maintaining the rest of the code unchanged.

\subsection{Reaction Pruning with Dependency Graphs}

In the situation when we already know which reactions were enabled at the previous time step, we are able to derive which reactions will surely not be enabled in the current time step. In this way, it would not be necessary to check whether all the reactants are present and all the inhibitors absent. This can be obtained by exploiting the \emph{dependency graph} of the reactions.

Let $V = A$ be the set of vertices of the graph, and let the set $E$ of edges be defined as all pairs of reactions $(r_i,r_j)$ such that the products of $r_i$ and the reactants of $r_j$ have a non-empty intersection. Intuitively, we have that reaction $r_i$ produces some of the reactants of $r_j$.

Let $T$ be the state of a reaction system. Let $H = \{r_{i_1}, r_{i_2}, \ldots, r_{i_h}\}$ be the set of reactions enabled in $T$, and let $K$ be the set defined as:
\begin{align*}
  K = \{ r_j \;|\; (r_i, r_j) \in E \text{ and } r_i \in H\}
\end{align*}
Then, in the absence of context, the reactions enabled in $\res_\rsA(T)$ will necessarily be a subset of $K$. This is due to the fact that the reactions enabled in $H$ only produced reactants for the reactions in $K$. Not all reactions in $K$ are necessarily enabled, since some reactants might be missing, or some inhibitors might be present. It is however important to stress that only the reactions in $K$ need to be checked, since no reaction outside it can be enabled.

\begin{example}
  If we draw the dependency graph for the reaction system $\rsA$ used in our working example, we obtain the following graph:
  \[
    \begin{tikzpicture}[->,>=stealth',shorten >=1pt,auto,node distance=2cm,semithick]
      \node[state] (r1) {$r_1$};
      \node[state] (r2) [above right of=r1] {$r_2$};
      \node[state] (r3) [below right of=r2] {$r_3$};
      
      \path (r1) edge[loop above] node {} (r1)
      edge node{} (r2)
      (r2) edge node{} (r3)
      (r3) edge node{} (r1);
    \end{tikzpicture}
  \]
  If we start with $T = \{b, d\}$, we register that only reaction $r_3$ is actually enabled in $T$. This means that at the next time step we only need to check whether reaction $r_1$ (the only one directly reachable from $r_3$) is actually enabled. In fact, we already know, from the properties of the dependency graph, that none of the reactants necessary for $r_2$ and $r_3$ were produced.
\end{example}

In our implementation in \clrs we use two additional data structures: a bit vector $v_A$ of length $|A|$ to memorize the reactions to be checked at each time step, and a vector $v_S$ of length $|S|$ in which each entry is a vector of indices of $v_A$. An element in position $i$ of $v_S$ identifies the reactions, represented as indices of $v_A$, that have the $i$-th entity as a reactant. In particular, at each time step when a reaction $j$ is enabled, all entries of $v_A$ that appear in $v_S[j]$ are set to one. So, when we check what reactions are enabled, only those with the corresponding bit of $v_A$ set to one are actually checked. This means that we only check reactions for which at least one reactant has been produced.

\subsection{Matrix-based Simulation}

Given a state $T \subseteq S$, let $\vec{t}$ be the column vector in $\{0,1\}^n$ representing the characteristic function of $T$ (i.e., $\vec{t}_i$ is $1$ when the $i$-th entity is present in $T$ and $0$ otherwise). Furthermore, for a set $B \subseteq A$ of reactions, let $\vec{b}$ be the row vector representing the characteristic function of $B$. The column vectors will be used to represent the current state of the reaction system, while the row vectors will represent the reactions enabled in the current state of the system.

Let $\matR$ be a matrix with $|A|$ rows and $|S|$ columns, where the entry $r_{i,j} = 1$ represents the fact that the $i$-th reaction in $A$ has the $j$-th entity of $S$ as a reactant and $r_{i,j} = 0$ otherwise. The matrix $\matR$ will be called the \emph{reactants matrix} in the rest of the paper. Together with $\matR$, we also define the \emph{inhibitors matrix} $\matI$, also having $|A|$ rows and $|S|$ columns. An entry $in_{i,j}$ of $\matI$ is $1$ when the $i$-th reaction in $|A|$ has the $j$-th entity as an inhibitor and $0$ otherwise. Similarly to $\matR$ and $\matI$, we can define the \emph{products matrix} $\matP$ as having $|A|$ rows and $|S|$ columns. Here, the entry $p_{i,j}$ is $1$ when the $j$-th entity is a product of the $i$-th reaction, and $0$ otherwise.

Furthermore, let $c_1 \,:\, \Z^{|A|} \to \{0,1\}^{|A|}$ be defined for each $\vec{x} = (x_0,\ldots, x_{|A|-1})$ as:
\begin{align*}
  & c_1(\vec{x})_i =
    \begin{cases}
      1 & \text{if $x_i \ge |R_i|$} \\
      0 & \text{otherwise}
    \end{cases} & \text{for $0 \le i < |A|$}
\end{align*}
where $R_i$ is the set of reactants of the $i$-th reaction in $A$. Similarly, let $c_2 \,:\, \Z^{|S|} \to \{0,1\}^{|S|}$ be the function that, given a vector $x$ of $|S|$ integers, returns a vector of the same length $|S|$, with $1$ in the positions corresponding to positive entries in $x$ an $0$ in all other entries.

Our first claim is that, given a state $T \subseteq S$, the vector $\vec{a}$ defined as:
\begin{align*}
  & \vec{a} = c_1(\vec{\hat{a}}) & \text{with } & \vec{\hat{a}} = (\matR - \matI) \vec{t}
\end{align*}
is the vector of reactions enabled in $T$. First of all, notice that the $i$-th entry of $\vec{\hat{a}}$ is defined as:
\begin{align*}
  & \hat{a}_i = \sum_{j=0}^{|S|-1} (r_{i,j} - in_{i,j}) t_j & \text{for $0 \le i < |A|$}
\end{align*}
since the $i$-th row of $\matR$, the number of ones is equal to $|R_i|$, the only possibility for $\vec{\hat{a}}_i$ to be at least $|R_i|$ is that the following two conditions are met:
\begin{align*}
  & \sum_{j=0}^{|S|-1} r_{i,j}t_j = |R_i| & \sum_{j=0}^{|S|-1} in_{i,j}t_j = 0
\end{align*}
The first condition takes into account the fact that exactly $|R_i|$ entries on the $i$-th row of $\matR$ are $1$ and all the others are $0$, hence the second condition must thus be met in order to have $\hat{a}_i \ge |R_i|$. It follows from the definition of $\matR$ and $\vec{t}$ that the first condition is met only when all the reactants of the $i$-th reaction of $A$ are present in $T$. Similarly, from the definition of $\matI$ and $\vec{t}$, it follows that the second condition holds only when no inhibitor of the $i$-th reaction of $A$ is present. Therefore, the vector $\vec{a}$ is actually the vector of the reactions enabled in $T$. We now claim that the vector $\vec{t'} \in \{0,1\}^{|S|}$ defined as follows:
\begin{align*}
  & \vec{t'} = c_2 \left( \vec{a} \; \matP \right)
\end{align*}
is actually the vector corresponding to $T' = \res_\rsA(T)$. In fact, an element $t'_i$ of $\vec{t'}$ is $1$ only when the following condition holds:
\begin{align*}
  & \sum_{j = 0}^{|A|-1} a_j p_{j, i} > 0
\end{align*}
which, by definition of $\vec{a}$ and $\matP$, is true only when \emph{at least} one enabled reaction produces the $i$-th entity in $|S|$. Therefore, the result function $\res_\rsA$ acting on a set $T \subseteq S$ can also be computed as:
\begin{align*}
 c_2 \big(  c_1\big( (\matR - \matI) \vec{t} \big) \matP \big)
\end{align*}
In the presence of a context $C$, denoted by the vector $\vec{c}$, the state on which $\res_\rsA$ must be computed, namely, $T \cup C$, can be computed as $\max(\vec{t}, \vec{c})$ where $\max$ is computed element-wise.

Therefore, the next state of a reaction system with or without context can be computed by two matrix-vector multiplications and three additional operations: namely, $c_1$, $c_2$, and optionally $\max$ (clearly, $\matR - \matI$ needs to be computed only once, so it is not counted here).

\begin{example}
  In our working example, the matrix $\matR - \matI$, the product matrix $\matP$, and the state $T$ of the system are encoded as follows:
  \begin{align*}
    & \matR - \matI = 
      \begin{bmatrix}
        1 & 1 & -1 & 0 \\
        1 & -1 & -1 & 0 \\
        0 & 0 & -1 & 1
      \end{bmatrix} &
      \matP =
      \begin{bmatrix}
        1 & 1 & 0 & 0 \\
        0 & 0 & 0 & 1 \\
        0 & 1 & 0 & 0
      \end{bmatrix} &&
      \vec{t} = \begin{bmatrix} 0 \\ 1 \\ 0 \\ 1 \end{bmatrix}
  \end{align*}
  The first operation $(\matR - \matI) \vec{t}$, to find which reactions are enabled, produces the following row vector:
  \begin{align*}
    & \begin{bmatrix}
      1 & 1 & -1 & 0 \\
      1 & -1 & -1 & 0 \\
      0 & 0 & -1 & 1 \\
    \end{bmatrix}
    \begin{bmatrix} 0 \\ 1 \\ 0 \\ 1 \end{bmatrix}
    = \begin{bmatrix}
      1 & -1 & 1
      \end{bmatrix}
  \end{align*}
  which is then ``normalized'' using the function $c_1$, considering that $|R_1| = 2$, $|R_2| = 1$, and $|R_3| = 1$:
  \begin{align*}
    c_1\left(\begin{bmatrix}1 & -1 & 1 \end{bmatrix}\right) =
    \begin{bmatrix} 0 & 0 & 1\end{bmatrix}.
  \end{align*}
  The resulting row vector is then multiplied by $\matP$ to obtain the following column vector:
  \begin{align*}
    & \begin{bmatrix} 0 & 0 & 1\end{bmatrix}
      \begin{bmatrix}
        1 & 1 & 0 & 0 \\
        0 & 0 & 0 & 1 \\
        0 & 1 & 0 & 0
      \end{bmatrix} =
      \begin{bmatrix}
        0 \\ 1 \\ 0 \\ 0
      \end{bmatrix}               
  \end{align*}
  which, after normalizing it with $c_2$ (which keeps it unchanged in this example), represents the new state vector of the system. By performing the same operations again we will obtain the null vector of four elements, which represents the next state of the reaction system.
\end{example}

\section{Experimental Settings}
\label{sec:settings}

In the experimental phase we compared \brsim, \heresy (both CPU and GPU, denoted here by \texttt{heresy} and \texttt{heresy-gpu}), \clrs, \clrs using the dependency graph (denoted by \texttt{cl-rs graph}), and a proof-of-concept implementation of the matrix-based simulation in Python 3 with Numpy (from now on denoted by \texttt{matrix}). All the tests were performed in a system with an Intel Core i7-3537U as a CPU, clocked at 2.5 GHz, and a Nvidia GeForce Titan X GPU, with 3584 cores clocked at 1417 MHz. The operating system was Ubuntu 16.04.5 with kernel 4.4.0. The Common Lisp compiler used was SBCL version 1.4.14, whereas the version of Python was 3.5.2 with Numpy 1.13.3. The results about the performances of \brsim and \heresy were taken from~\cite{Nobile2017a}, since the machine where the tests were performed is the same. All tests for every combination of the parameters and for the \emph{ErbB} real-world model were repeated $30$ times to obtain realistic average running times.

\subsection{Synthetic Models}

As a first case, we employed the same synthetic models (i.e., not actually created to model any real-world phenomenon) used in~\cite{Nobile2017a}. Each reaction system is generated according to three parameters:
\begin{itemize}
  \item The number of entities, that is, the size of the background set.
  \item The number of reactions.
  \item A parameter $\alpha \in [0,1]$ used to control the ``size'' of the reactions. In particular, for each reaction three numbers, $r$, $i$, and $p$ are extracted according to a binomial distribution with parameters $|S|$ and $\alpha$. With some additional checks to ensure that the reactions created were actually valid (see~\cite{Nobile2017a} for the details), those three numbers are used as the number of reactants, inhibitors, and products of the reaction.
\end{itemize}
A set of parameters is denoted by $|S|\times|A|\times\alpha$ (e.g., $1000 \times 1000 \times 0.1$).
We used four combinations of number of entities and reactions, each of them associated with three values for the parameter $\alpha$:
\begin{itemize}
  \item \emph{small systems}: $10 \times 10 \times \alpha$ with $\alpha \in \{0.01, 0.05, 0.1\}$;
  \item \emph{medium systems}: $100 \times 100 \times \alpha$ with $\alpha \in \{0.01, 0.05, 0.1\}$;
  \item \emph{large systems}: $1000 \times 1000 \times \alpha$ and $2000 \times 2000 \times \alpha$ with $\alpha \in \{0.01, 0.05, 0.1\}$.
\end{itemize}
Furthermore, each reaction system also had an associated context sequence consisting of $1000$ steps, each step independently generated by uniformly sampling the background set. Therefore, each simulation was performed for $1000$ time steps, irrespective of the number of reactions and entities.

\subsection{ErbB Model}

In addition to the synthetic models, we also employed a real-world model of the ErbB receptor signal transduction in human mammary epithelial cells, as described in~\cite{helikar2013comprehensive}. The original model was not in the form of a reaction system and consisted of $245$ nodes and $1100$ arcs, representing, respectively, heterogeneous cellular components and the biochemical reactions among them. In~\cite{Nobile2017a} the model was converted to a reaction system with $6720$ reactions involving $246$ entities. A context sequence of length $1000$ was also generated as described in~\cite{helikar2013comprehensive}.

\section{Experimental Results}
\label{sec:results}

\begin{figure}
  \centering
  \includegraphics[width=0.45\textwidth]{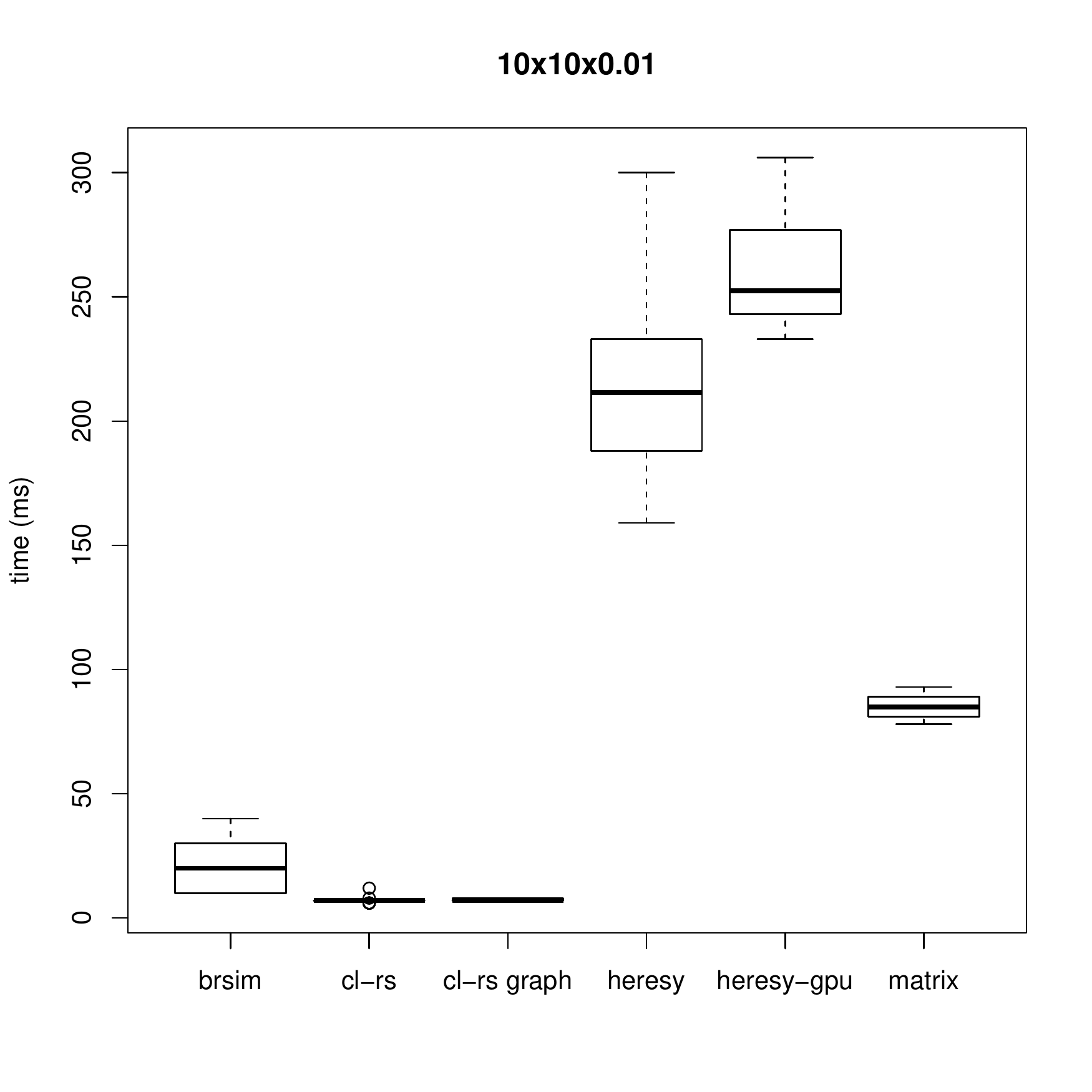}
  \includegraphics[width=0.42\textwidth]{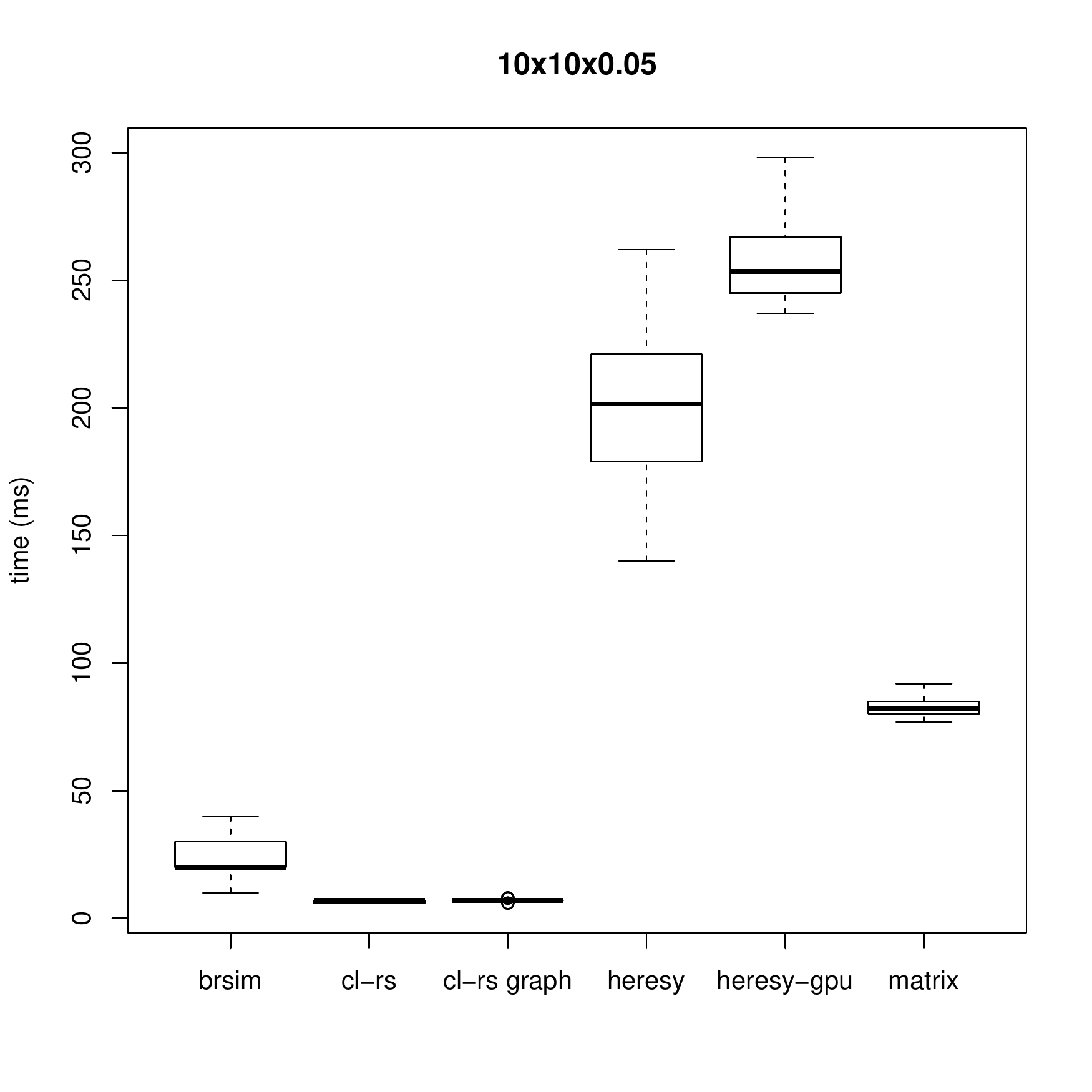}
  \includegraphics[width=0.42\textwidth]{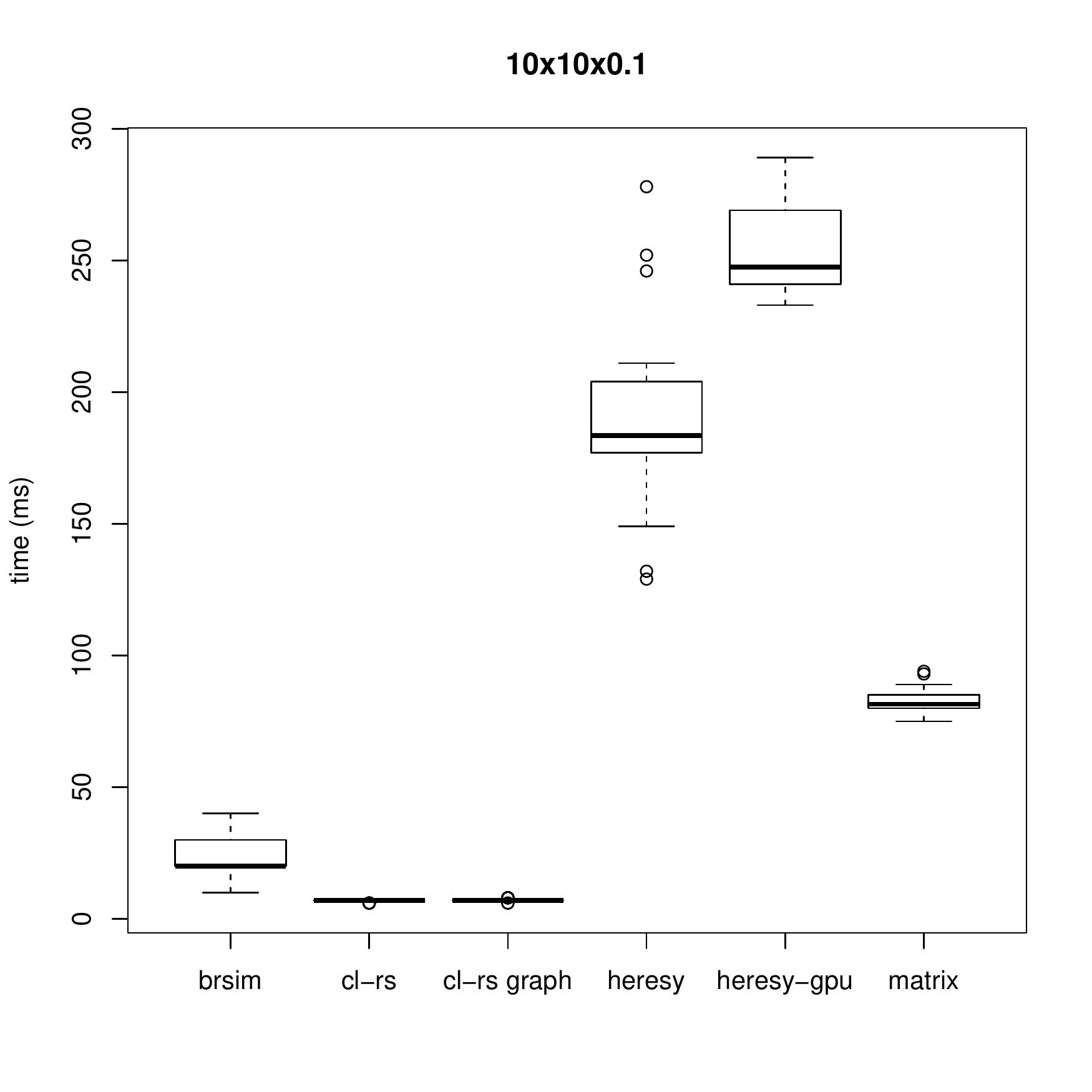}
  \includegraphics[width=0.42\textwidth]{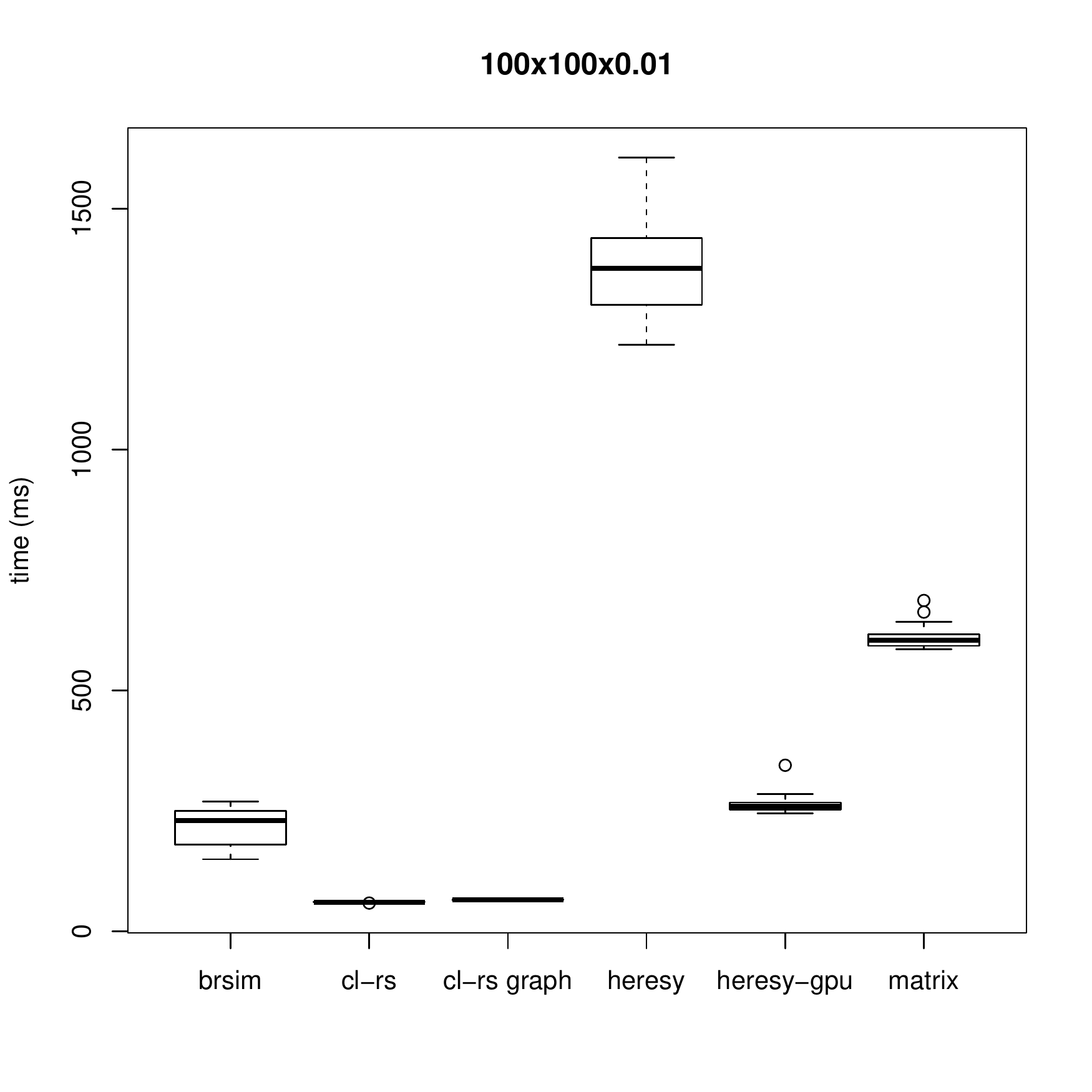}
  \includegraphics[width=0.42\textwidth]{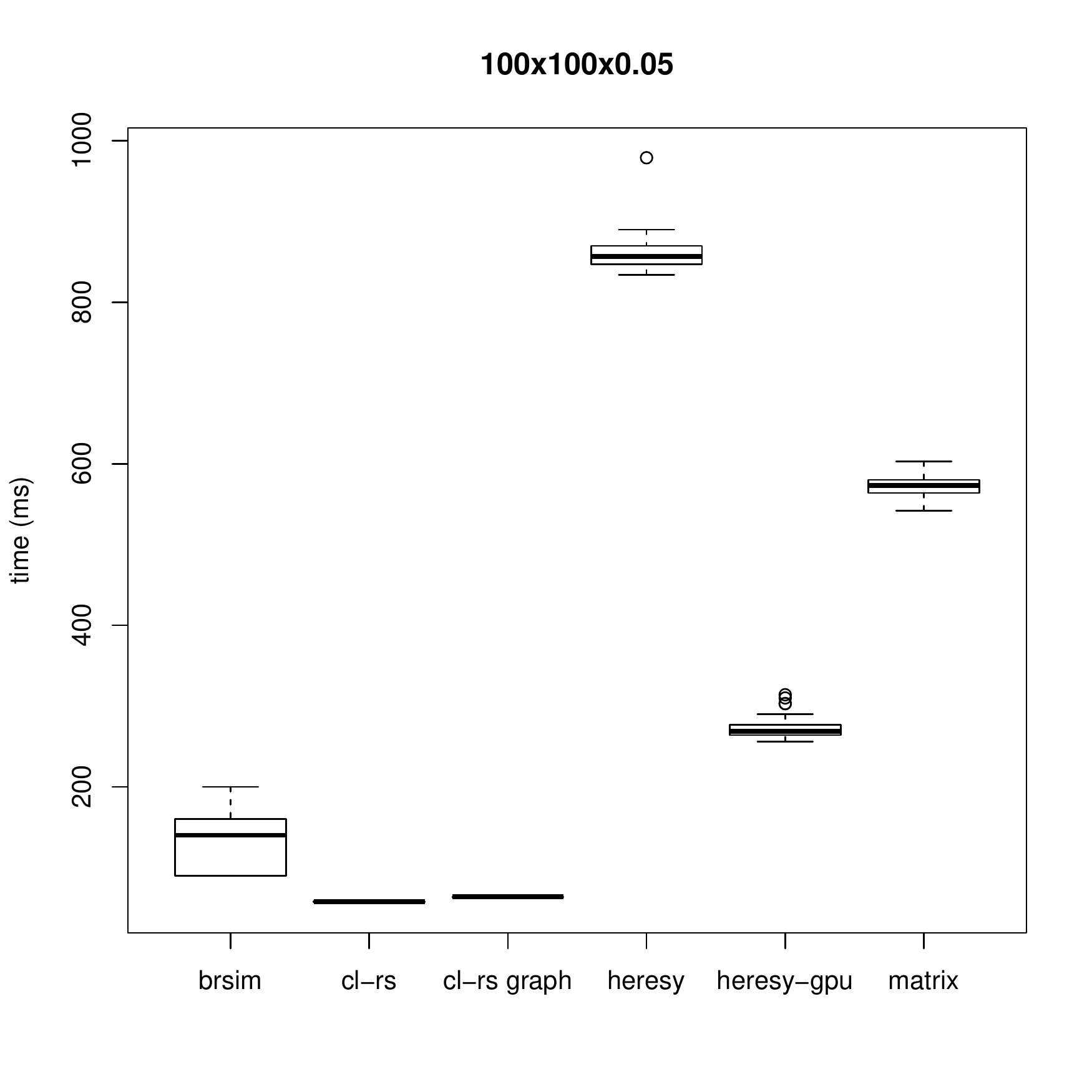}
  \includegraphics[width=0.42\textwidth]{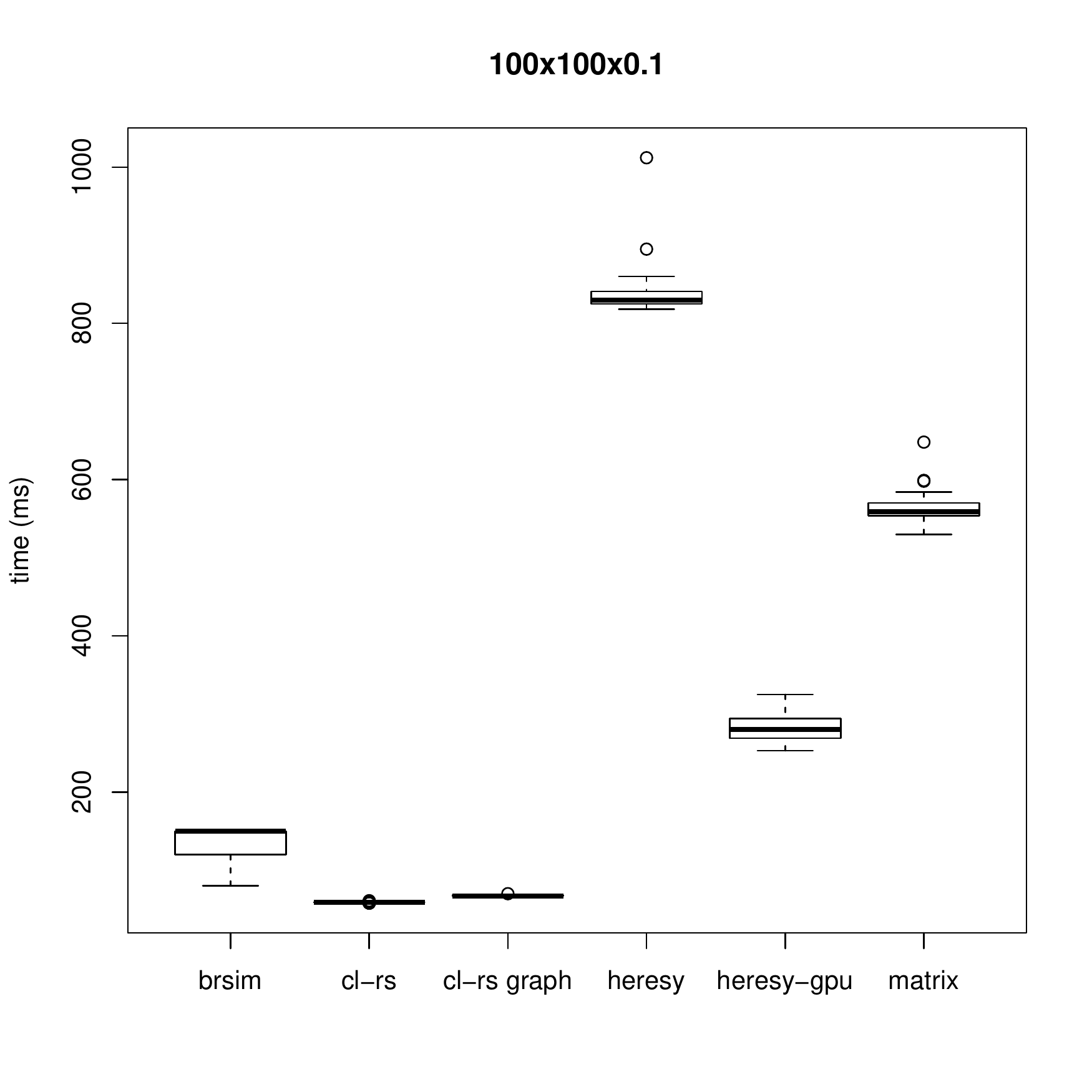}
  \caption{\label{fig:size-10-100} Results for small and medium systems. From left to right and top to bottom: $\alpha = 0.01$, $\alpha = 0.05$, and $\alpha = 0.1$ for $10 \times 10 \times \alpha$ systems, and $\alpha = 0.01$, $\alpha = 0.05$, and $\alpha = 0.1$ for $100 \times 100 \times \alpha$ systems.}
\end{figure}

\begin{figure}
  \centering
  \includegraphics[width=0.42\textwidth]{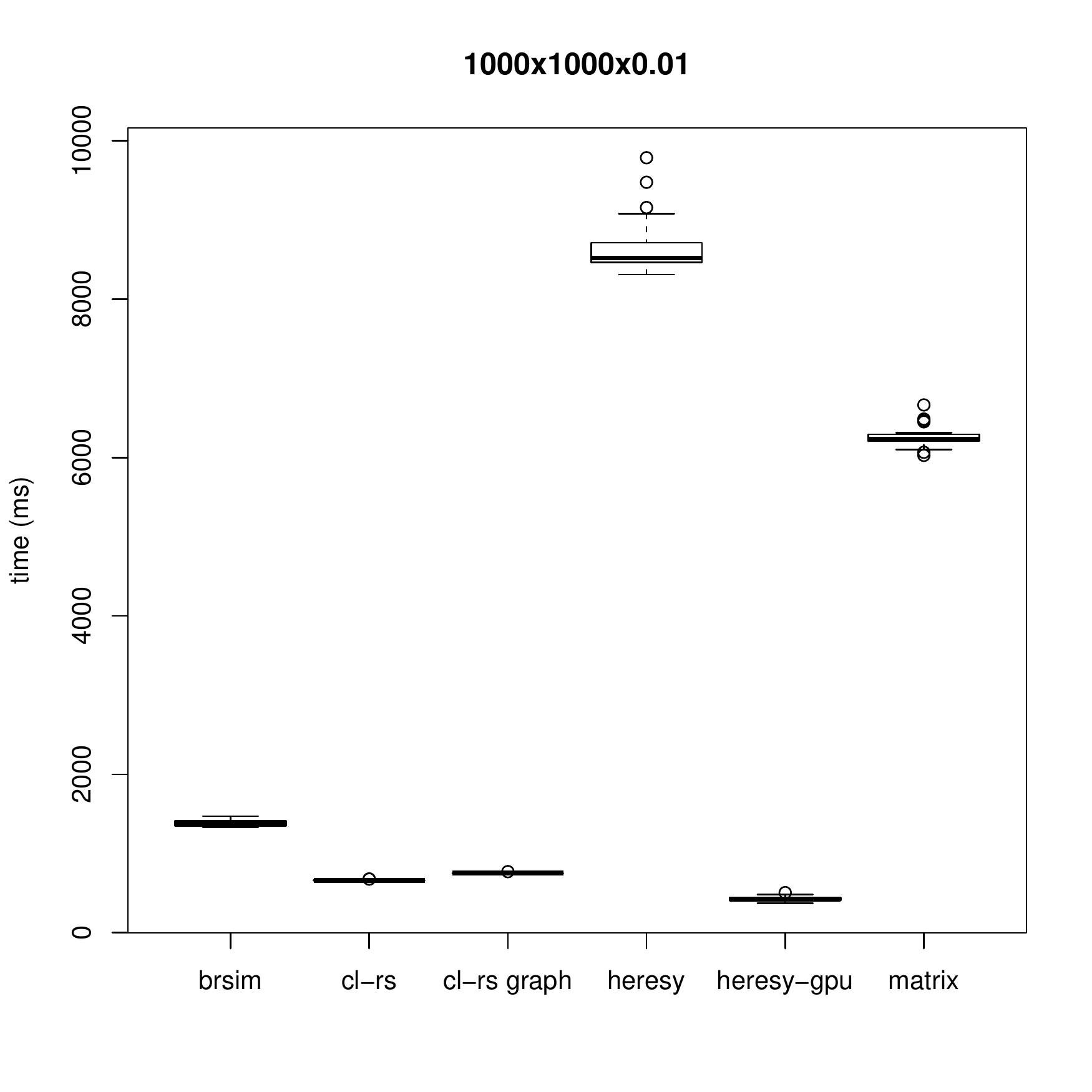}
  \includegraphics[width=0.42\textwidth]{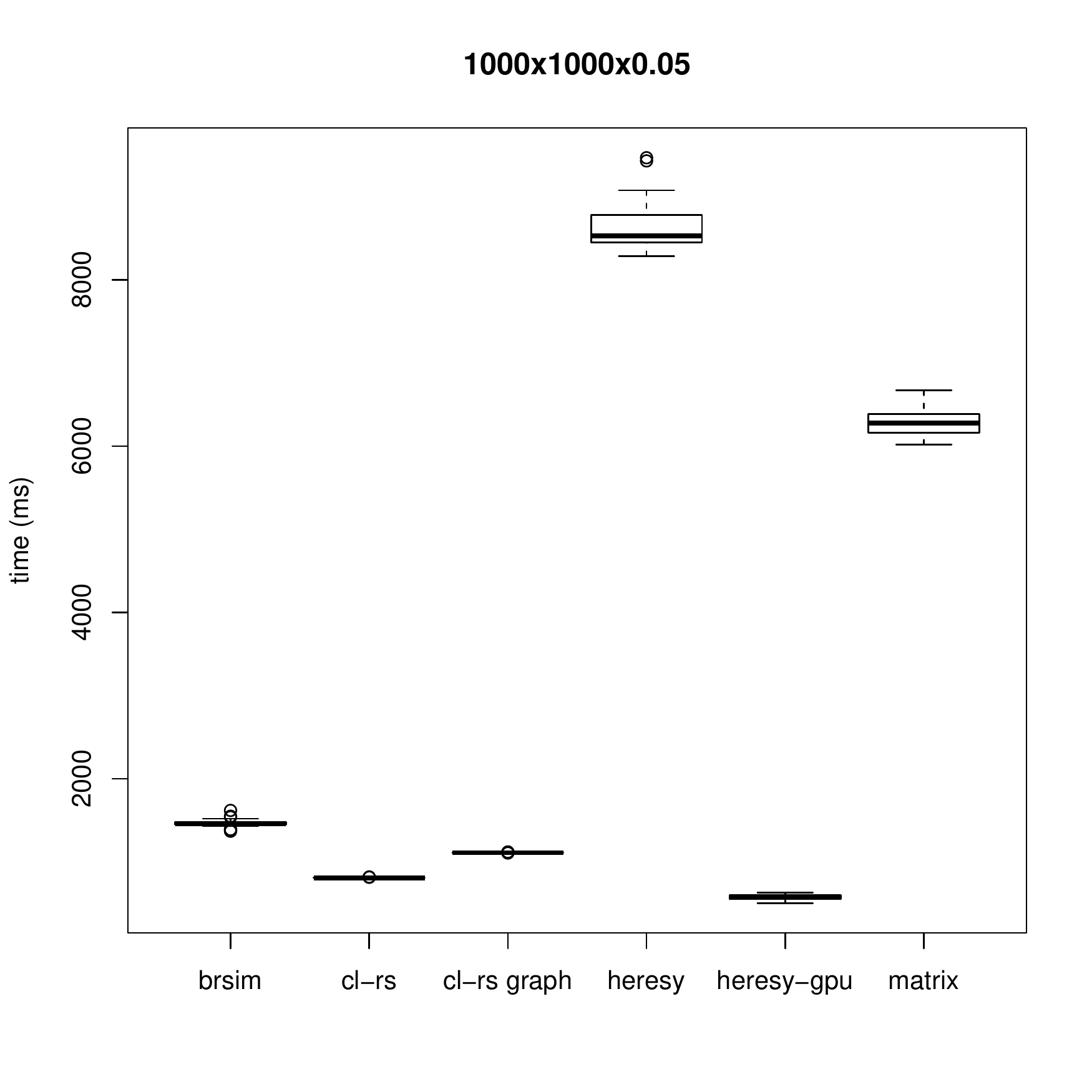}
  \includegraphics[width=0.42\textwidth]{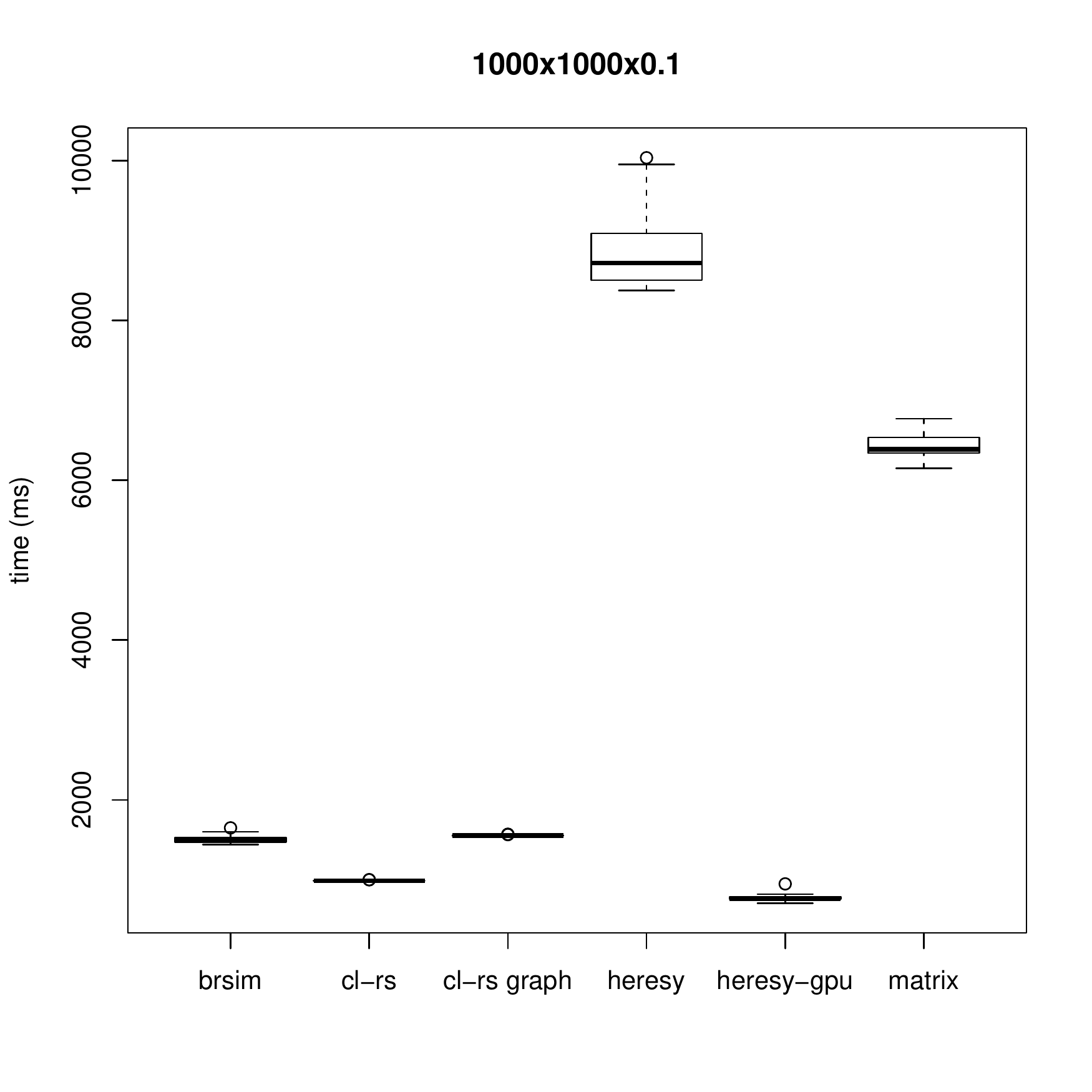}
  \includegraphics[width=0.42\textwidth]{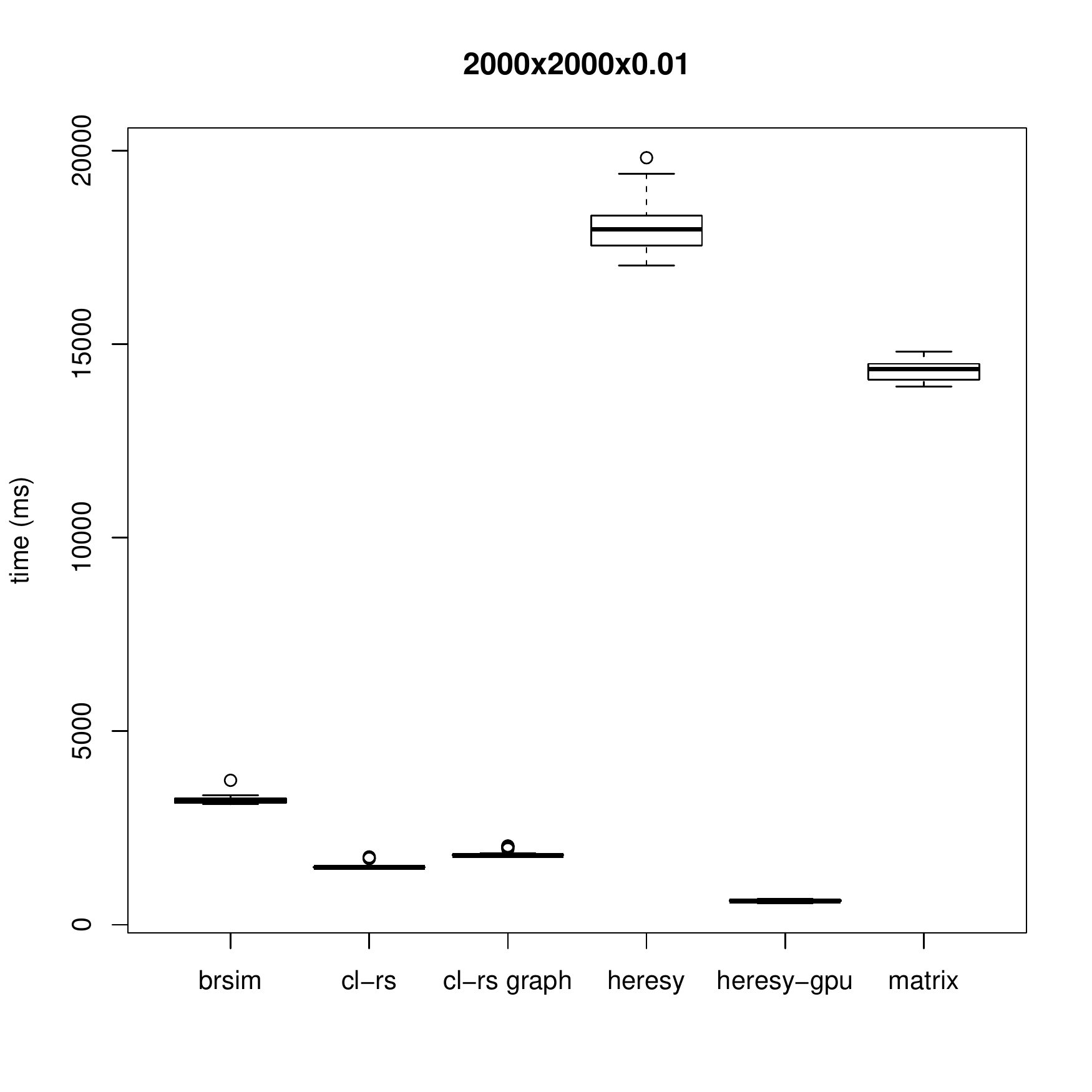}
  \includegraphics[width=0.42\textwidth]{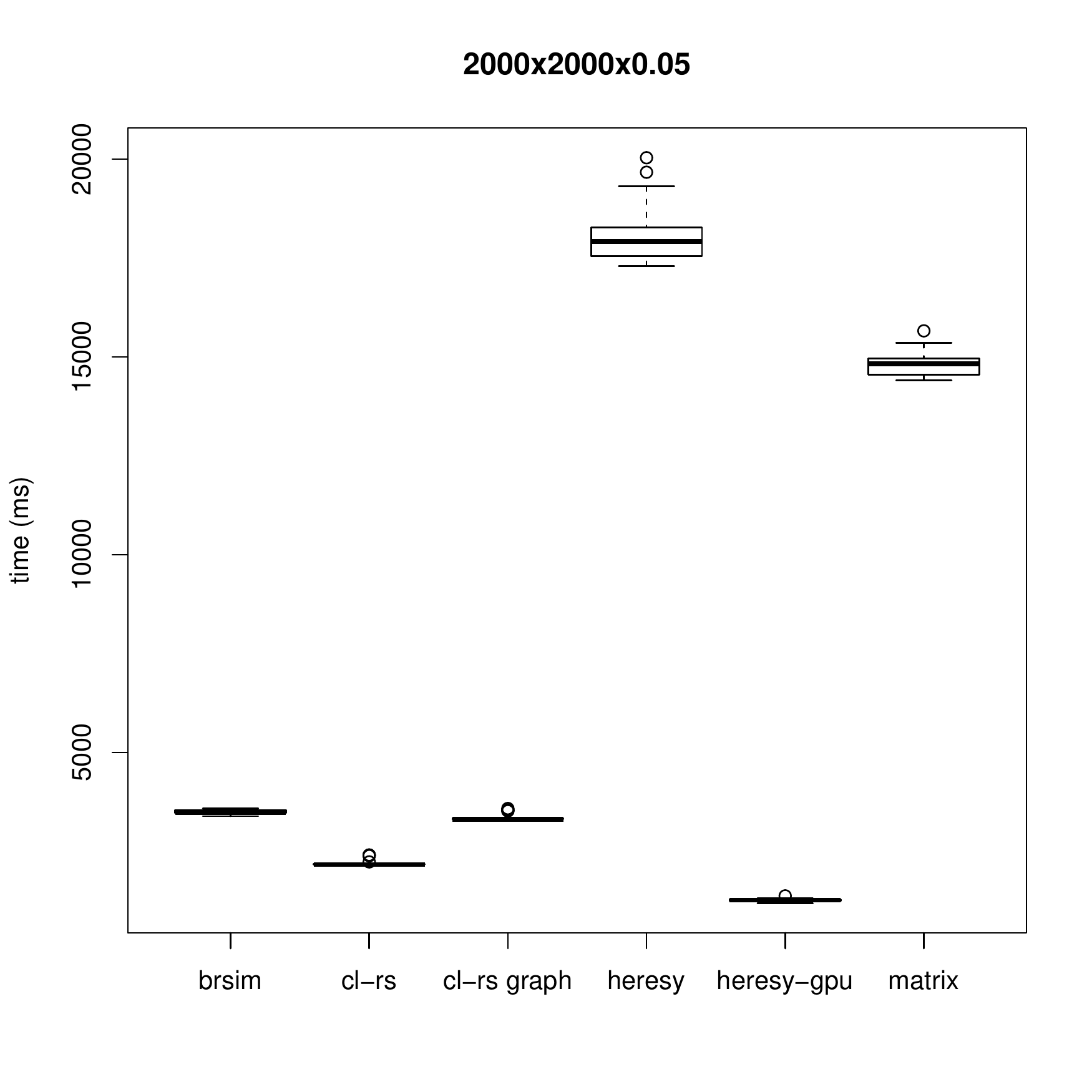}
  \includegraphics[width=0.42\textwidth]{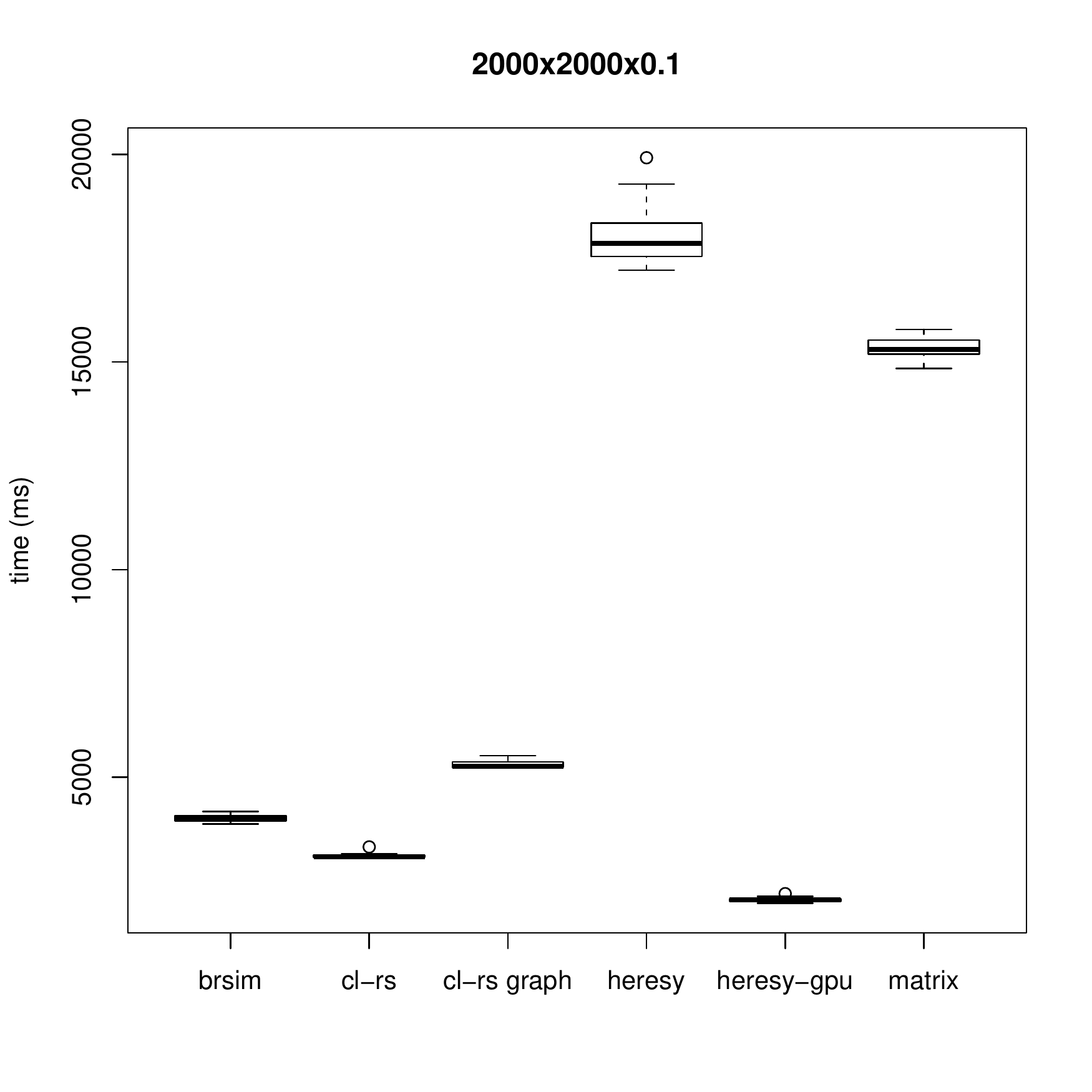}
  \caption{\label{fig:size-1000-2000} Results for large systems. From left to right and top to bottom: $\alpha = 0.01$, $\alpha = 0.05$, and $\alpha = 0.1$ for $1000 \times 1000 \times \alpha$ systems, and $\alpha = 0.01$, $\alpha = 0.05$, and $\alpha = 0.1$ for $2000 \times 2000 \times \alpha$ systems.}
\end{figure}

\begin{figure}
  \centering
  \includegraphics[width=0.42\textwidth]{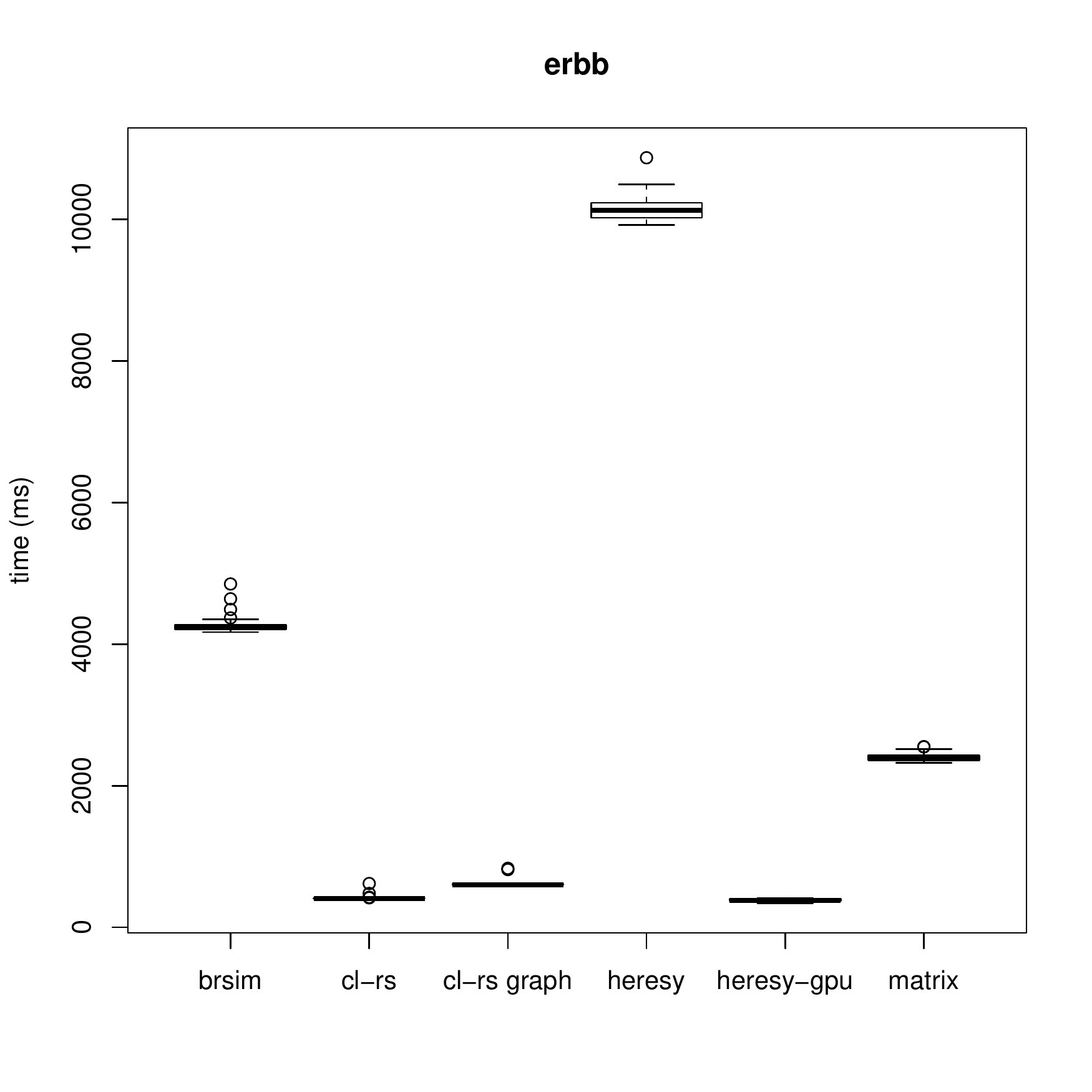}
  \includegraphics[width=0.42\textwidth]{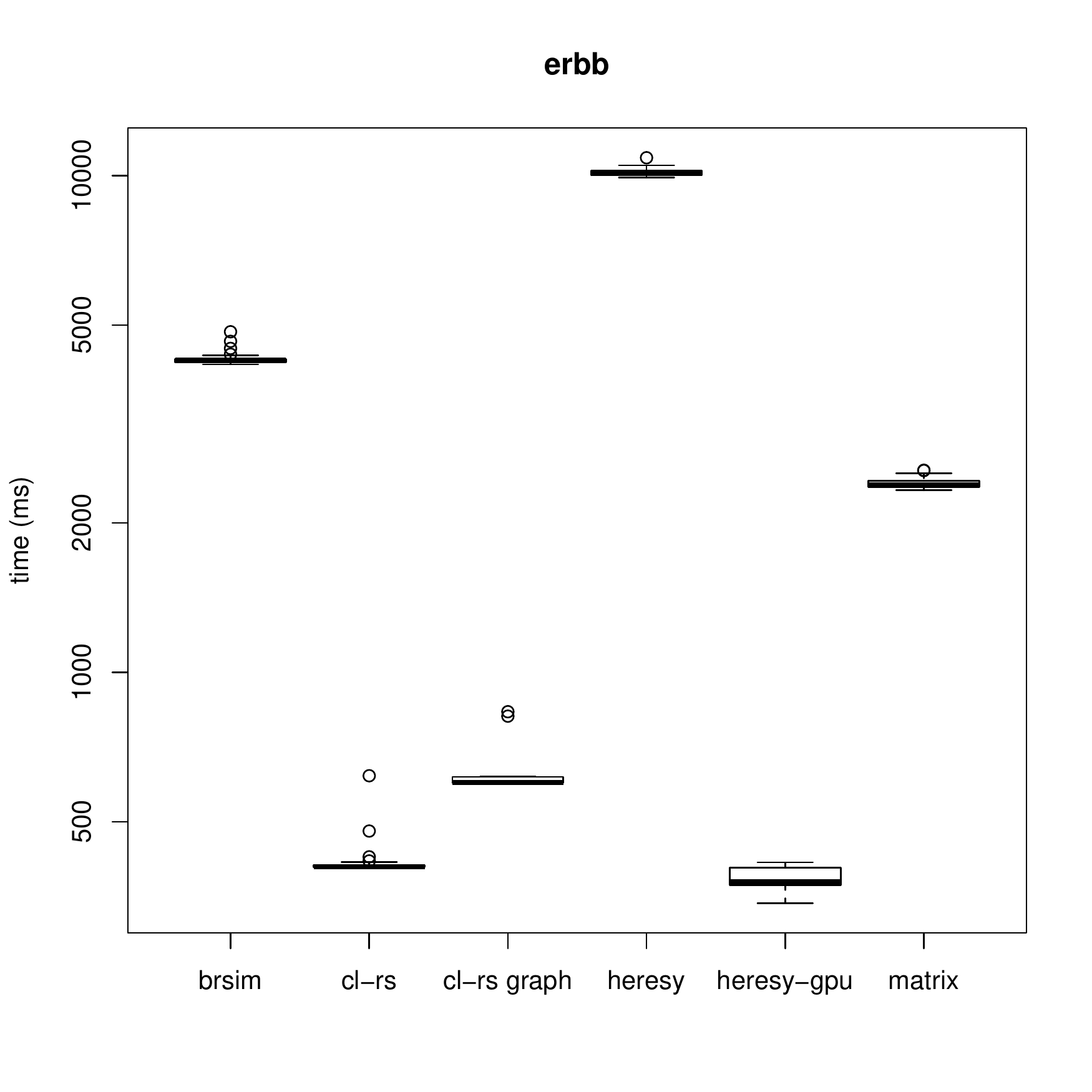}
  \caption{\label{fig:ErbB} The results on the ErbB model (on the left). The plot on the right reports the same results but using a logarithmic scale for the ordinate axis, in order to make the results for \clrs, \texttt{cl-rs grah}, and \texttt{heresy-gpu} more readable.}
\end{figure}

The boxplots of the running times for the different experiments are presented in Figure~\ref{fig:size-10-100} for small and medium  systems, in Figure~\ref{fig:size-1000-2000} for large systems, and in Figure~\ref{fig:ErbB} for the ErbB model. A summary of the average running times of the simulators on the different models is presented in Table~\ref{tab:results}.

From the results on small systems it is possible to observe that both \clrs and \texttt{cl-rs graph} are the fastest systems, with an average runtime of $7$ milliseconds, about one third of the time required by \brsim and from $26$ to  $29$ times faster than \texttt{heresy}. The \texttt{matrix} program is more than three times slower than \brsim, but still faster than \heresy.

For medium systems, the two fastest programs are still \clrs and \texttt{cl-rs graph}, the first requiring about $60$ms to perform each simulation, and the second from $64$ to $68$ms. This is about twice faster than \brsim and significantly faster than \texttt{heresy} on both CPU and GPU. \texttt{matrix} is slower than \brsim, but still faster than \texttt{heresy}.

For large systems the situation changes significantly. The parallelism of the GPU-based simulator is able to offset the cost incurred by moving data to and from the main memory, and thus \texttt{heresy-gpu} is able to obtain the best performances. The fastest CPU-based program remains \clrs, with \texttt{cl-rs graph} being slower than \brsim for systems with $2000$ entities and reactions with a large amount of reactants (i.e., $\alpha = 0.1$); in the other cases it remains faster than \brsim. The \texttt{matrix} method is still faster than \texttt{heresy} on CPU, but slower than \brsim.

Finally, for the real-world system, ErbB, both \texttt{heresy-gpu} and \clrs perform similarly, in the order of $400$ms. \texttt{cl-rs graph} is slower by about $200$ms. The \texttt{matrix} program is almost twice faster than \brsim ($2.40$s vs $4.28$s).

\begin{table}
  \centering
  \small
  \setlength{\tabcolsep}{5pt}
  \begin{tabular}{lcccccc}
    \toprule
    & \textbf{brsim} & \textbf{cl-rs} & \textbf{cl-rs graph} & \textbf{heresy} & \textbf{heresy-gpu} & \textbf{matrix} \\
    \midrule
    \emph{$10 \times 10 \times 0.01$}     & 23 (10)    & 7 (1)     & 7 (0)     & 209 (32)    & 260 (21)  & 85 (5)      \\
    \emph{$10 \times 10 \times 0.05$}     & 23 (9)     & 7 (0)     & 7 (1)     & 201 (30)    & 259 (18)  & 83 (4)      \\
    \emph{$10 \times 10 \times 0.1$}      & 21 (8)     & 7 (0)     & 7 (1)     & 188 (32)    & 255 (17)  & 82 (5)      \\
    \emph{$100 \times 100 \times 0.01$}   & 217 (39)   & 61 (1)    & 66 (1)    & 1382 (95)   & 264 (19)  & 610 (23)    \\
    \emph{$100 \times 100 \times 0.05$}   & 128 (34)   & 58 (1)    & 64 (1)    & 862 (26)    & 274 (16)  & 573 (14)    \\
    \emph{$100 \times 100 \times 0.1$}    & 129 (28)   & 59 (1)    & 68 (1)    & 840 (36)    & 285 (20)  & 565 (22)    \\
    \emph{$1000 \times 1000 \times 0.01$} & 1383 (37)  & 662 (6)   & 752 (5)   & 8643 (344)  & 426 (30)  & 6254 (130)  \\
    \emph{$1000 \times 1000 \times 0.05$} & 1468 (51)  & 810 (3)   & 1113 (2)  & 8632 (298)  & 574 (32)  & 6295 (166)  \\
    \emph{$1000 \times 1000 \times 0.1$}  & 1506 (44)  & 989 (4)   & 1554 (6)  & 8857 (455)  & 776 (43)  & 6435 (151)  \\
    \emph{$2000 \times 2000 \times 0.01$} & 3218 (119) & 1508 (75) & 1828 (82) & 18024 (686) & 617 (29)  & 14316 (248) \\
    \emph{$2000 \times 2000 \times 0.05$} & 3501 (59)  & 2203 (81) & 3355 (88) & 18106 (703) & 1268 (35) & 14835 (312) \\
    \emph{$2000 \times 2000 \times 0.1$}  & 4014 (76)  & 3104 (49) & 5315 (85) & 18014 (640) & 2050 (51) & 15330 (238) \\
    \emph{ErbB}                         & 4285 (143) & 417 (41)  & 618 (56)  & 10152 (190) & 386 (20)  & 2400 (59)   \\
    \bottomrule
\end{tabular}
\caption{\label{tab:results} A summary of the running times in the different test problems. The first number represents the average runtime in milliseconds, while the second number (in parentheses) is the standard deviation, also in milliseconds.}
\end{table}

It is interesting to discuss the possible reasons for the observed performances. Starting with \clrs, its use of compact data structures seems to allow most of the important data to fit into the fast processor cache. This would explain why \clrs can be less than two times faster than \brsim (e.g., for $2000$ reactions and entities with $\alpha=0.1$) and some other times more than $10$ times faster (e.g., for ErbB). A similar explanation also works for \texttt{cl-rs graph} which, however, has to manage additional data structures and, therefore, its performances are more influenced by the type of reactions in the system. It would be interesting to see for which kinds of reaction systems the dependency graph-based approach would work better. Of particular interests are the performances of the \texttt{matrix} program. The synthetic systems studied are all such that the matrices $\matR - \matI$ and $\matP$ are square matrices. This appears to have a negative influence, since in the case of ErbB, where the two matrices are rectangular with $6720$ rows and $246$ columns, the performances are almost two times better than \brsim. It remains open to see if an optimized version of the algorithm would be able to be the best performer on CPU or if a GPU implementation could be faster than \texttt{heresy-gpu}.

\section{Conclusions and Future Works}
\label{sec:conclusion}

In this paper we have introduced new approaches to the simulation of reaction systems, mainly the ability to ``prune'' reactions using a dependency graph, and the possibility of modeling the next-state function of a reaction system as operations on matrices and vectors together with some additional ``clipping'' steps. We have introduced a new CPU-based simulator, \clrs, that proved to be faster than all currently existing CPU-based simulators and able to attain performances similar to the existing GPU-based simulator in a large scale real-world model.

There are still many possible avenues for additional research. For example, an in-depth complexity analysis of the different algorithms for different classes of reaction systems (e.g., with ``short'' reactions, with different proportions of reactions and entities, etc.) is still missing. This would also help in finding for which classes of reaction systems the dependency graph-based approach can be a sensible choice. Furthermore, in the dependency graph approach, there are possible variations to consider. We have studied the case for a \emph{positive} dependency graph, that is, where the edges are defined by looking at the reactants. We can also try to employ a \emph{negative} dependency graph, where the edges are defined by looking at the inhibitors. This would allow to immediately exclude reactions whose inhibitors were produced in the previous step. It would also be possible to combine the two approaches to further limit the set of reactions to be checked at every time step. Finally, the matrix-based approach was tested with only a proof-of-concept code. It would be interesting to optimize it using ad-hoc methods and to implement it on a GPU.

\bibliographystyle{fundam}
\bibliography{Bibliography}

\end{document}